\documentclass[aps,prb,superscriptaddress,twocolumn]{revtex4}
\usepackage{amssymb}
\usepackage{graphicx}
\usepackage{amsmath}
\usepackage[colorlinks=true,linkcolor=blue,citecolor=blue,urlcolor=blue]{hyperref}

\newcommand{\be}{\begin{equation}}
\newcommand{\ee}{\end{equation}}

\newcommand{\bs}{\begin{split}}
\newcommand{\es}{\end{split}}

\usepackage[normalem]{ulem}
\usepackage{amsmath,amssymb}
\usepackage{multirow}
\usepackage{xcolor,soul}
\usepackage{srcltx}
\usepackage{hyperref,graphicx}

\begin{document}

\title{Magnetic Control of the Non-Hermitian Skin Effect in Two-Dimensional Lattices}
\author{Stefano Longhi}
\thanks{stefano.longhi@polimi.it}
\affiliation{Dipartimento di Fisica, Politecnico di Milano, Piazza L. da Vinci 32, I-20133 Milano, Italy}
\affiliation{IFISC (UIB-CSIC), Instituto de Fisica Interdisciplinar y Sistemas Complejos, E-07122 Palma de Mallorca, Spain}

\begin{abstract}
The non-Hermitian skin effect (NHSE) --- the anomalous boundary accumulation of an extensive number of bulk modes --- has emerged as a hallmark of non-Hermitian physics, with broad implications for transport, sensing, and topological classification. A central open question is how magnetic or synthetic gauge fields influence this boundary phenomenon. Here, we develop a theoretical framework for magnetic control of the NHSE along line boundaries in two-dimensional single-band lattices. Using a non-Hermitian extension of the anisotropic Harper--Hofstadter model as a representative example, we show that magnetic fields suppress the geometric skin effect in reciprocal models, whereas skin localization can persist in nonreciprocal systems. The analysis disentangles the interplay of flux, nonreciprocity, and boundary geometry, revealing that magnetic fields mitigate or suppress the NHSE through distinct physical mechanisms --- such as bulk localization via Landau or Anderson physics, or the restoration of effective reciprocity. In particular, the geometry-dependent skin effect in reciprocal systems is found to be fragile against even weak magnetic fields.
\end{abstract}
\maketitle

\section{Introduction}
The non-Hermitian skin effect (NHSE) -- the accumulation of an extensive number of bulk eigenstates at system boundaries -- has emerged as a defining feature of non-Hermitian systems \cite{R5,R6,R7,R8,R9,R10,R11,R11b,Ru1,Ru2,R11c,R12,R13,R14,R15,R16,R17,R17b,R17c,R18,R19,R20,Hughes2021,R21,R21b,R21c,R21d,R22,R23,R24,R25,R26,R27,R28,R28b,R28c,R29,R29b,R30,R30b,R30c,R30d,R31}, with broad implications for condensed matter, photonics, acoustics, and engineered quantum platforms (for reviews, see \cite{R20,R21c,R22,R23,R26,R27,R28,R29b,R31}). The NHSE reflects the breakdown of conventional bulk-boundary correspondence \cite{R5,R7,R9} and underlies unconventional transport, enhanced sensing, and novel topological classifications. It was first identified in one-dimensional models with nonreciprocal (asymmetric) hopping \cite{R5,R6,R7,R9}, such as the Hatano-Nelson chain \cite{Hatano1996,Hatano1997,Longhi2015,Gong2018}, and has since been generalized to higher-dimensional systems, reciprocal lattices, many-body, incoherent and nonlinear models (see, e.g., \cite{R21,R21b,R28b,WW1,WW2,WW3,WW4,WW5,WW6,YY1,YY2,YY3,YY4,YY5,YY6,YY7,YY8,YY9,YY10,YY11,YY12,YY13,YY14,YY15,YY15b,YY16,YY17,YY18,YY19,YY20,YY21,YY22,YY23,YY24,YY25} and references therein). Several theoretical frameworks -- including non-Bloch band theory, biorthogonal bulk-boundary correspondence, and real-space topological invariants -- have provided a unified understanding of the NHSE by connecting spectral winding, boundary accumulation, and topological protection \cite{R20,Hughes2021,R21c,R22,R23,R26,R27,R28,Gong2018}. Experimental demonstrations in photonic, acoustic, mechanical, topolectrical, and ultracold atomic platforms \cite{R15,R16,R17,R18,R19,R21b,R21d,R25,R28b,R28c,E1,E2,E3} have further established its generality and tunability.

Magnetic or synthetic gauge fields offer a natural mechanism to control the NHSE \cite{M1,M2,M3,M4,M5,M6,M7,M8,M9,M10,M11,M12}. Their effects, however, are highly system dependent, particularly in two-dimensional lattices where distinct types of NHSE and boundary dependencies arise \cite{R21}. In nonreciprocal systems, magnetic fields can mitigate or suppress first-order NHSE by restoring Landau-level-like bulk states and shrinking the skin topological area, thereby reducing boundary accumulation \cite{M1,M2,M9}. Similar suppression has been observed for pseudomagnetic fields engineered from spatially inhomogeneous gauge configurations \cite{M8}. Conversely, magnetic fields can enhance or induce skin localization in other regimes: they stabilize second-order skin modes by protecting line gaps in the complex spectrum \cite{M5}, or drive field-induced transitions in spinful non-Hermitian systems \cite{M10,M11,M12}. Moreover, the suppression of skin modes observed for weak magnetic fields and low-energy regimes associated with Landau localization \cite{M1} can decrease an disappears as the magnetic flux is increased \cite{M4}. These contrasting behaviors illustrate the complex interplay between gauge fields and non-Hermitian topology, which remains far from fully understood. In particular, the distinctive role of gauge fields on skin states in nonreciprocal versus reciprocal models -- the latter exhibiting the so-called geometric NHSE \cite{R21,R28b,R28c} -- remains poorly characterized.

In this work, we present a theoretical framework for analyzing the magnetic control of the NHSE in two-dimensional single-band lattices with strip geometries. Using a non-Hermitian extension of the anisotropic Harper--Hofstadter model~\cite{M8,HH1,HH2} as a representative example, we examine the interplay among flux, nonreciprocity, and boundary geometry. The results show that magnetic fields suppress the geometry-dependent skin effect in reciprocal systems, rendering it fragile even under weak flux, while skin localization in nonreciprocal systems is not universally suppressed. In the latter case, gauge-field-induced mixing of boundary modes qualitatively modifies the localization pattern and can mitigate, or in some regimes suppress, skin accumulation through Landau- or Anderson-type bulk localization. By contrast, in reciprocal systems the suppression of the geometry-dependent skin effect arises primarily from the field-induced restoration of effective reciprocity rather than bulk localization.

This study provides a coherent interpretation of how gauge fields influence boundary accumulation in reciprocal and nonreciprocal two-dimensional models, highlighting the fragility of the geometric NHSE in the presence of magnetic flux. The analysis builds upon and connects previous results across condensed-matter, photonic, and acoustic settings, contributing to the broader understanding of boundary phenomena in non-Hermitian physics.

\section{Model and basic equations}

We consider a single-band two-dimensional lattice with one atom per unit cell. The wavefunction is denoted by $\psi(\mathbf{R})$, where $\mathbf{R}$ labels the lattice sites. The sites are defined as
\begin{equation}
\mathbf{R} = n \mathbf{a}_X + m \mathbf{a}_Y = n a_X \mathbf{u}_X + m a_Y \mathbf{u}_Y, \quad n,m \in \mathbb{Z},
\end{equation}
where $\mathbf{a}_X = a_X \mathbf{u}_X$ and $\mathbf{a}_Y = a_Y \mathbf{u}_Y$ are the primitive vectors of the Bravais lattice $\mathcal{B}$, forming an angle $\alpha$ with each other (Fig.1), and 
$\mathbf{u}_{X,Y}$ are the associated unit vectors.
The system is subject to a uniform magnetic field $B$ perpendicular to the crystal $(X,Y)$ plane. We aim to investigate how the magnetic field controls NHSE localization toward a {\em rather arbitrary} edge direction $x$ (first-order NHSE), forming an angle $\theta$ with the principal direction $X$. We remark that the direction $x$ does not necessarily coincide with the $X$ or $Y$ axes of the primitive vectors. To this end, we consider a {\em strip geometry}: the lattice is infinite along the $x$ direction (with periodic boundary conditions, $x$-PBC) and finite along $y$ (with open boundary conditions, $y$-OBC). Eventually, $y$-PBC condition can be assumed along the $y$ direction as well to relate the emergence of the NHSE toward the $y$ edges, under $y-$OBC, with the point-gap topology of the reduced Hamiltonian, under  $y$-PBC.  
The $x$ axis is chosen such that a subset of sites of the Bravais lattice lies along it, as illustrated schematically in Fig.1. This condition is met provided that the ratio
\begin{equation}
\frac{a_X \sin \theta}{a_Y \sin ( \alpha- \theta)} \equiv \frac{p}{q}
\end{equation}
is rational, with $p,q$ coprime integer numbers.  The $y$ axis is rotated relative to $x$ by the same angle $\alpha$ between the primitive vectors.  

\begin{figure}[t]
 \centering
    \includegraphics[width=0.48\textwidth]{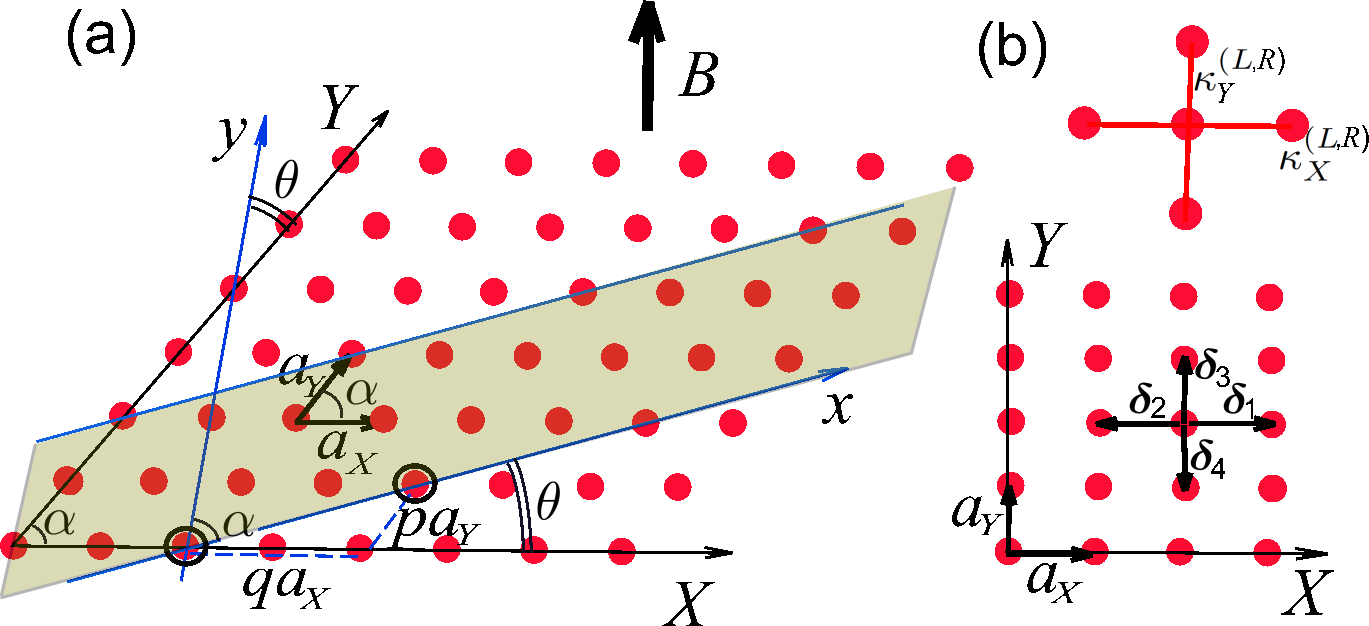}
   \caption{(a) Schematic of a 2D crystal with a magnetic field $B$ applied orthogonal to the crystal plane $(X,Y)$. $\mathbf{a}_X = a_X \mathbf{u}_X$ and $\mathbf{a}_Y = a_Y \mathbf{u}_Y$ are the primitive vectors of the Bravais lattice $\mathcal{B}$, forming an angle $\alpha$ with each other. We consider a strip geometry (shaded area) defined along the $x$ and $y$ directions, where the system is infinite along $x$ (equivalently, periodic boundary conditions are imposed, $x$-PBC) and finite along $y$ with open boundary conditions ($y$-OBC). The $y$ axis is rotated relative to $x$ by the same angle $\alpha$ between the primitive vectors, while the $x$ axis is rotated by $\theta$ with respect to the principal axis $X$ of the crystal. (b) Schematic of the non-Hermitian anisotropic Harper-Hofstadter  model (rectangular lattice with nearest-neighbor hopping).}
    \label{fig1}
\end{figure}

To describe the effect of the magnetic field on the energy spectrum and eigenstate localization in the strip geometry, we adopt the Landau gauge in the $(x,y)$ frame:
\begin{equation}
\mathbf{A}(\mathbf{r}) = (B \sin \alpha \, y, 0,0),
\end{equation}
where the obliquity factor $\sin \alpha$ in Eq.(3) accounts for the non-orthogonality of the $x$ and $y$ directions. The tight-binding eigenvalue equation reads
\begin{equation}
E \psi(\mathbf{R}) = \sum_{\boldsymbol{\delta}} t(\boldsymbol{\delta}) \, \exp\left(i \phi_{\mathbf{R}, \mathbf{R}+\boldsymbol{\delta}} \right) \psi(\mathbf{R} + \boldsymbol{\delta}),
\end{equation}
where $\boldsymbol{\delta} \in \mathcal{B}$, $t(\boldsymbol{\delta})$ is the hopping rate between sites ${\mathbf{R}}$ and $\mathbf{R}+ \boldsymbol{\delta}$, and
\begin{equation}
\phi_{\mathbf{R}, \mathbf{R}+ \boldsymbol{\delta}}= 2 \pi \int_{\mathbf{R}}^{\mathbf{R}+\boldsymbol{\delta}} d \mathbf{r} \cdot \mathbf{A}(\mathbf{r})
\end{equation}
is the Peierls' phase (gauge phase) contribution to the hopping amplitude arising from the vector potential (we assume $h=e=1$, i.e. a unit flux quantum). The integral on the right hand side of Eq.(5) is taken along the straight line connecting $\mathbf{R}$ and $\mathbf{R}+\boldsymbol{\delta}$. We indicate by $\mathcal{B}_H \in \mathcal{B}$ the subset of vectors of the Bravais lattice $\mathcal{B}$ such that $t(\boldsymbol{\delta}) \neq 0$, i.e. for which particle hopping can arise from site $\mathbf{R}+\boldsymbol{\delta}$ to site $\mathbf{R}$. The lattice is {\em reciprocal} provided that, for any
$\boldsymbol{\delta} \in \mathcal{B}_{H}$, then also $-\boldsymbol{\delta} \in \mathcal{B}_{H}$ and 
\begin{equation}
 |t(-\boldsymbol{\delta})|=|t(\boldsymbol{\delta})|.
\end{equation}
If the additional condition $t(-\boldsymbol{\delta})=t^*(\boldsymbol{\delta})$ holds, than the lattice is also {\em Hermitian}.\\
 To solve the eigenvalue equation (4) in the strip geometry, it is worth assuming the variable $\mathbf{R}=(x,y)$ as a continuous one. Owing to the Landau gauge (3), the Peierls' phase $\phi_{\mathbf{R}, \mathbf{R}+ \boldsymbol{\delta}}$ turns out to be a function of $y$ solely, namely one has
 \begin{equation}
  \phi_{\mathbf{R}, \mathbf{R}+ \boldsymbol{\delta}}= 2 \pi B y  \sin \alpha (\delta_x+ \delta_y \cos \alpha) + \theta_{\boldsymbol{\delta}}
 \end{equation}
where we have set
 \begin{equation}
 \theta_{\boldsymbol{\delta}}= \pi B \delta_y   \sin \alpha  (\delta_x+ \delta_y \cos \alpha)
 \end{equation}
 and ${\boldsymbol{\delta}}=\delta_x \mathbf{u}_x+\delta_y \mathbf{u}_y$. Here, $\mathbf{u}_{x,y}$ are the unit vectors of the rotated directions $x$ and $y$ (Fig.1). Since the Peierls' phase depends only on \(y\), under $x-$PBC we consider wavefunctions of the form
\begin{equation}
\psi(x,y) = \phi(y) e^{i k_x x},
\end{equation}
with \( k_x \) a constant quasi-momentum along the \(x\)-direction. Substitution of Eq.(9) into Eq.(4) yields the following difference equation for $\phi(y)$
\begin{equation}
E \phi(y)= \sum_{ \delta_y} \tau_{\delta_y} (y)  \phi(y+\delta_{y})
\end{equation}
where we have set
\begin{eqnarray}
\tau_{\delta_y}(y) & = & \sum_{\delta_x}  t ( {\boldsymbol{\delta}})  \times \\
& \times & \exp \left\{ 2 \pi i B y \sin \alpha (\delta_x+ \delta_y \cos \alpha) +ik_x \delta_x+i \theta_{\boldsymbol{\delta}}   \right\}. \nonumber
\end{eqnarray}
Since the vector ${\boldsymbol{\delta}} \in \mathcal{B}_H$ belongs to the Bravais lattice,  there are two integers $n_X$ and $n_Y$ such that ${\boldsymbol{\delta}}=n_Xa_X \mathbf{u}_X+n_Y a_Y \mathbf{u}_Y$, so that after some straighforward algebra one obtains
\begin{eqnarray}
\delta_x & = & n_Xa_X \frac{\sin(\alpha+\theta)}{\sin \alpha}+n_Ya_Y \frac{\sin \theta}{\sin \alpha}\\
\delta_y & = & - n_Xa_X \frac{\sin \theta}{\sin \alpha}+n_Ya_Y \frac{\sin (\alpha-\theta)}{\sin \alpha}.
\end{eqnarray}
From Eqs.(2) and (13), one can write
\begin{equation}
\delta_y=a(-n_Xp+n_Yq)
\end{equation}
with 
\[ a \equiv a_X \frac{\sin \theta }{p \sin \alpha}. \]
This means that the spatial shifts $\delta_y$, as $\boldsymbol{\delta}$ varies in $\mathcal{B}_H$, is an integer multiple than $a$, and thus Eq.(10) is formally analogous to the eigenvalue problem on a 1D lattice with lattice constant $a$. After letting $y=na$ and $\phi_n=\phi(y=na)$, Eq.(10) reads
\begin{equation}
E \phi_n=\sum_l \tau _{l}(n) \phi_{n+l}  \equiv \sum_m {\mathcal{H}}_{n,m} \phi_m
\end{equation}
where
\begin{eqnarray}
\tau_{l}(n) & = & \sum_{  \boldsymbol{\delta} | \delta_y=l a } t( \boldsymbol{\delta})  \times \\
& \times &  \exp \left\{ 2 \pi i B n a  \sin \alpha (\delta_x+ la \cos \alpha) +ik_x \delta_x+i \theta_{\boldsymbol{\delta}}   \right\} \nonumber
\end{eqnarray}
is the effective hopping rates between sites distant $la$. 
Equations (15) and (16) represent an effective 1D lattice model in the $y$ direction of the slab, described by the spatial Hamiltonian $\mathcal{H}$, and provide the starting point to investigate the skin localization of bulk modes toward the $x$-edge direction and the role of the magnetic field $B$ in controlling skin localization.

\section{Magnetic control of non-Hermitian skin effect}
Let us first consider the case $B=0$, corresponding to the vanishing of Peierls' phases. The Bloch energy spectrum of the 2D lattice, under $X$-PBC and $Y$-PBC, is given by
\begin{equation}
E=H(\mathbf{k})=\sum_{\boldsymbol{\delta} \in \mathcal{B}_N} t(\boldsymbol{\delta}) \exp(i \mathbf{k} \cdot \boldsymbol{\delta} ).
\end{equation}  
The area covered by $H(\mathbf{k})$ on the complex plane, as $\mathbf{k}$ spans the first Brillouin zone,  is called
the spectral area. A general result on the universality of the NHSE, stated in Ref.\cite{R21}, is that the skin effect emerges under open boundary of generic geometry if and only if the
spectral area is nonzero. This universal form of the skin effect can be partitioned into two classes: the nonreciprocal skin effect and the generalized reciprocal skin effect. The former arises quite generally in systems with nonreciprocal hopping and is characterized by nonvanishing currents, whereas the latter occurs in reciprocal models, with the geometry-dependent skin effect \cite{R21} being a characteristic example. This special type of generalized reciprocal NHSE exhibits the unique feature that there always exists at least one boundary geometry for which the skin effect is completely absent.  
\par
In the effective 1D model obtained in the strip geometry of Fig.1(a), in the absence of the magnetic field ($B=0$)
the hopping rates $\tau_l(n)$ do not depend on $n$ and the effective 1D lattice defined by Eq.(15) displays discrete spatial invariance along $y$. The condition for the appearance of the NHSE under $y-$OBC, i.e. the localization of a macroscopic number of bulk modes toward the $y$ edges of the strip, is that the energy spectrum $E$ of Eq.(15), under $y-$PBC, shows a point-gap topology, i.e. it does not describe an open arc in complex energy plane. After letting $\phi_n=\exp(ik_y a n)$, the $y$-PBC energy spectrum  reads
\begin{equation}
H_y(k_y)= \sum_l \tau_l \exp(ik_yal).
\end{equation}
Therefore the NHSE under $y$-OBC appears if and only if the curve $H_x(k_y)$ describes a closed loop with point gap topology, i.e. if for some base energy $E_B$ the winding number \cite{R12,R20,R23,R26,R27,Gong2018}
\begin{equation}
w(E_B)=\frac{1}{2 \pi i} \int_0^{2 \pi/a}d k_y \frac{d}{dk_y} \log \left\{ H_y(k_y)-E_B \right\}
\end{equation}
is nonvanishing. According to the general results of Ref.\cite{R21}, for generic edge directions $y$ the winding number is non-vanishing, for both  reciprocal and non-reciprocal non-Hermitian lattices. A special form of reciprocal NHSE is the geometry-dependent skin effect that arises in certain reciprocal models: while the winding number $w$ is  non-vanishing for rather arbitrary $y$ direction, and thus accumulation of skin modes is observed under $y-$OBC, for special directions $y$, corresponding to special symmetries of the model (such as mirror symmetries), the winding number vanishes and the skin effect is not observed \cite{R21}. \par

When the magnetic field is turned on ($B \neq 0$), Eq.~(16) shows that the hopping amplitudes $\tau_l(n)$ (for $l \neq 0$) and the on-site energy $\tau_0(n)$ in the effective one-dimensional Hamiltonian $\mathcal{H}$ [Eq.~(15)] become spatially inhomogeneous. As a result, discrete translational symmetry is generally broken: the magnetic field effectively introduces an incommensurate disorder into the lattice, which can strongly affect skin localization of the eigenstates of $\mathcal{H}$ under $y$-OBC. Owing to the breakdown of translational invariance, various forms of winding numbers defined in real space have been proposed to characterize the emergence of skin states under $y$-OBC \cite{Gong2018,Hughes2021,Sarkar2022}. For instance, for a given base energy $E_B$, one may consider the polar decomposition of the shifted Hamiltonian $(\mathcal{H}-E_B) = \mathcal{Q}\mathcal{P}$, where the unitary factor $\mathcal{Q}$ captures the spectral winding in real space and thus provides a direct diagnostic of the NHSE.  
The winding number $w(E_B)$ can be introduced as \cite{Hughes2021,Sarkar2022}
\begin{equation}
w(E_B)= \frac{1}{L} {\rm Tr}  \left( \mathcal{Q}^{\dag } \left[ \mathcal{Q}, \mathcal{X}  \right]  \right)
\end{equation}
where $L$ is the lattice size and $\mathcal{X}$ is the position operator. 
Such a definition of the winding number $w(E_B)$ reduces to Eq.~(19) in the translationally invariant limit $B=0$. 
Another way to verify the suppression of the skin effect via winding numbers is to introduce an additional gauge phase $\Phi$ in the $y$-PBC and evaluate the winding number $w(E_B)$ as \cite{Gong2018}
\[
w(E_B)=\frac{1}{2 \pi i} \int_0^{2 \pi} d \Phi \, \frac{d}{d \Phi} \ln \det (\mathcal{H}-E_B),
\]
which is quantized, with $w(E_B)=0$ signaling the disappearance of the NHSE due to bulk localization of the eigenstates. In practice, numerical detection of the NHSE is often more straightforward by comparing the spectra under $y$-OBC and $y$-PBC. In the absence of the NHSE, the two spectra coincide in the thermodynamic limit ($L \to \infty$), apart from a finite, $L$-independent set of edge states under $y$-OBC. In the Hermitian limit, these boundary states correspond to the familiar chiral edge modes connecting bands with different Chern numbers. By contrast, in the presence of the NHSE, the spectra under $y$-OBC and $y$-PBC are markedly different. In the following, the appearance of the NHSE will be probed using this spectral comparison, together with the mean spatial distribution of eigenvectors of $\mathcal{H}$ under $y$-OBC.  

To summarize, the above analysis indicates that magnetic fields can generally control the skin effect, in close analogy with lattice models subject to incommensurate disorder \cite{M9}. This argument, however, does not resolve the distinct responses of reciprocal and nonreciprocal models to magnetic fields, nor does it clarify the fate of the geometry-dependent skin effect in reciprocal lattices. To disentangle these mechanisms in a controlled setting, the next section introduces a non-Hermitian extension of the anisotropic Harper--Hofstadter model \cite{HH1,HH2}, which provides a minimal and paradigmatic framework for exploring magnetic control of the NHSE in two-dimensional quantum Hall systems.  

\begin{figure*}[t]
 \centering
    \includegraphics[width=0.9\textwidth]{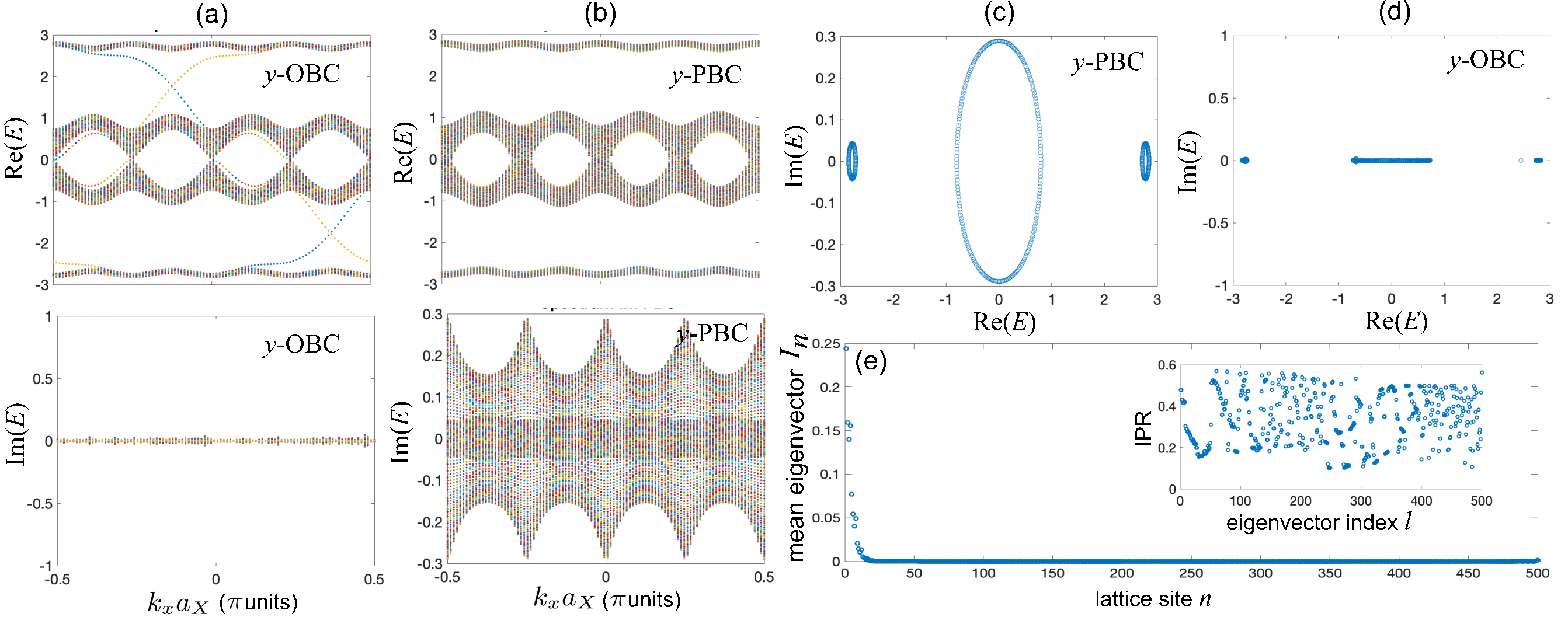}
   \caption{Effect of a rational magnetic flux on the energy spectrum and skin localization for the non-reciprocal Harper-Hofstadter model in a slab geometry [Eq.(32)]. Parameter values  are $h_X=0$, $h_Y=0.2$, $J_X=J_Y=1$ and magnetic flux $ \Phi=\alpha_x=\pi/2$. Lattice size in the $y$ direction is $L=500$. (a,b) Energy spectrum versus quasi-momentum $k_xa_X$ under $y$-OBC [panels (a)] and $y$-PBC [panels (b)]. Real and imaginary parts of $E$ are shown in the upper and lower panels, respectively. The curves in the gaps in panel (a) correspond to non-Hermitian extension of usual chiral edge states in the 2D quantum Hall model. (c,d) Energy spectrum in complex energy plane for $k_xa_X=0$ and for $y$-PBC [panel (c)] and $y$-OBC [panel (d)]. (e) Mean eigenvector distribution $I_n$ under $y$-OBC, clearly showing the persistence of the NHSE. The inset in (e) shows the IPR of the $L$ eigenstates under $y-$OBC.}
    \label{fig2}
\end{figure*}

\section{Magnetic control of the skin effect in the non-Hermitian Harper-Hofstadter model}
We consider a non-Hermitian extension of the Harper--Hofstadter model \cite{M8,HH1,HH2}, a paradigmatic model of two-dimensional quantum Hall physics \cite{HH1}. This framework is directly relevant to a variety of experimental platforms, including photonic crystals, topolectric circuits, and ultracold atoms, where synthetic magnetic fields and nonreciprocal couplings can be engineered. Specifically, we focus on a rectangular lattice ($\alpha=\pi/2$) with generally anisotropic and nonreciprocal nearest-neighbor hopping, defined by the set  
$ \mathcal{B}_H=\{\boldsymbol{\delta}_1, \boldsymbol{\delta}_2, \boldsymbol{\delta}_3, \boldsymbol{\delta}_4\}$,
with
\begin{equation}
\boldsymbol{\delta}_1=a_X \mathbf{u}_X, \; \boldsymbol{\delta}_2=-\boldsymbol{\delta}_1 \; ,  \boldsymbol{\delta}_3=a_Y \mathbf{u}_Y, \; \boldsymbol{\delta}_4=-\boldsymbol{\delta}_3
\end{equation}
and
\begin{equation}
t(\boldsymbol{\delta}_1)=\kappa_X^{(L)} \; , t(\boldsymbol{\delta}_2)=\kappa_X^{(R)}, t(\boldsymbol{\delta}_3)=\kappa_Y^{(L)} \; , t(\boldsymbol{\delta}_4)=\kappa_Y^{(R)}
\end{equation}
[see Fig.1(b)]. The rotation angle $\theta$ of the $x$ direction with respect to the principal crystal axis $X$ is assumed to satisfy the constraint (2), i.e.
\begin{equation}
{\rm tan} \theta= \frac{p}{q} \frac{a_Y}{a_X}
\end{equation} 
where $p$, $q$ are two irreducible (coprime) integers. After letting $a= (a_X/p)  \sin \theta= (a_Y /q) \cos \theta$, the effective one-dimensional eigenvalue equation in the $y$ direction is obtained by specializing Eqs.(15) and (16) using Eqs.(21) and (22). One obtains 
\begin{equation}
E \phi_n= \tau_p(n) \phi_{n+p}+ \tau_q(n) \phi_{n+q}+\tau_{-p} (n) \phi_{n-p}+\tau_{-q}(n) \phi_{n-q}
\end{equation}
where we have set
\begin{eqnarray*}
\tau_p(n) & = & \kappa_X^{(R)} \exp \left( -2 \pi i \alpha_x n +i \sigma_p  \right) \\
\tau_{-p}(n) & = & \kappa_X^{(L)} \exp \left( 2 \pi i \alpha_x n +i \sigma_{-p}  \right) \\
\tau_{q}(n) & = & \kappa_Y^{(L)} \exp \left( 2 \pi i \alpha_y n +i \sigma_q  \right) \\
\tau_{-q}(n) & = & \kappa_Y^{(R)} \exp \left( -2 \pi i \alpha_y n +i \sigma_{-q}  \right)
\end{eqnarray*}
and
\begin{eqnarray}
\alpha_x & = & Baa_X \cos \theta \; , \; \alpha_y= B a a_Y \sin \theta \\ 
\sigma_p & = & - k_xa_X \cos \theta-  \pi B a_X a p \cos \theta \\
\sigma_{-p} & = &  k_xa_X \cos \theta-  \pi B a_X a p \cos \theta \\
\sigma_q & = &  k_xa_Y \sin \theta+  \pi B a_Y a q \sin \theta \\
\sigma_{-q} & = &  - k_xa_Y \sin \theta+  \pi B a_Y a q \sin \theta 
\end{eqnarray}
To illustrate the above framework, two representative cases are analyzed in detail, after which a general discussion of the results is presented.  

\begin{figure*}[t]
 \centering
    \includegraphics[width=0.9\textwidth]{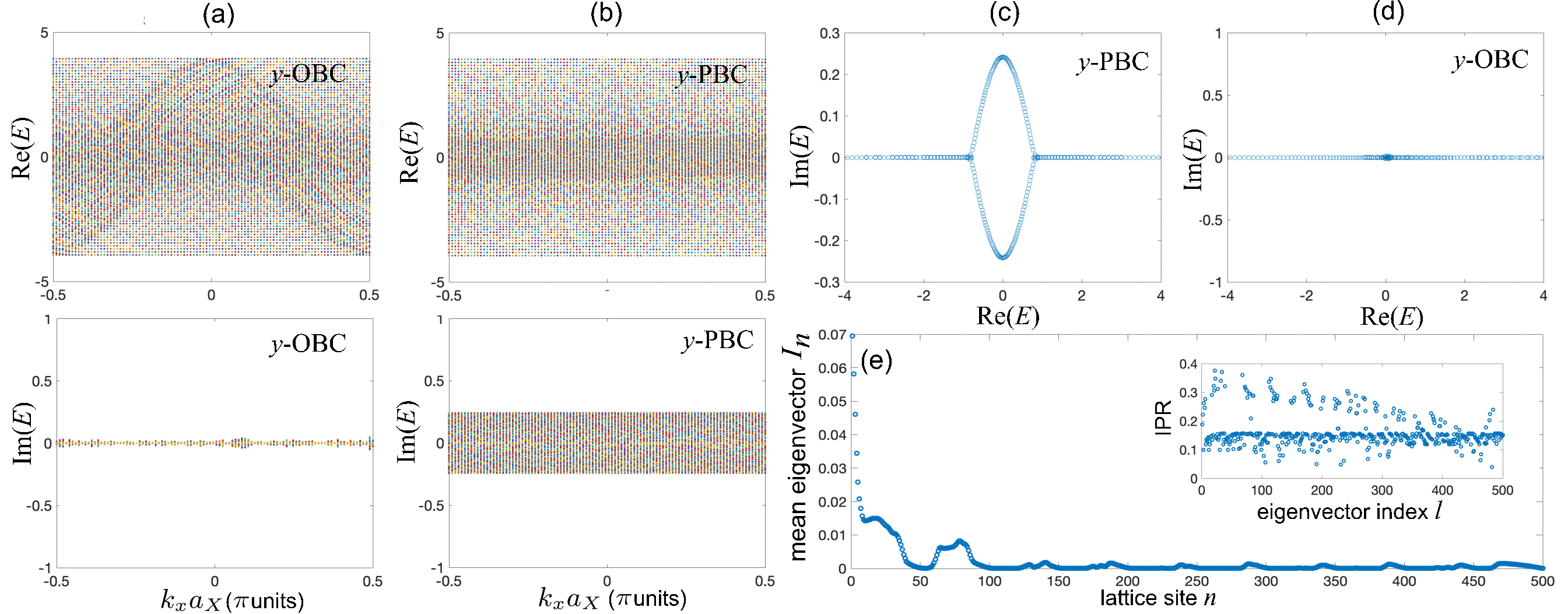}
   \caption{Same as Fig.2, but for a weak magnetic flux $\Phi=\pi/50$. Note that the skin area in panel (c) is reduced as compared to the case of Fig.2(c), and the corresponding mean eigenvector distribution $I_n$ in (e) tends to spread toward the bulk owing to the competing Landau localization.}
    \label{fig3}
\end{figure*}

\subsection{Non-reciprocal lattices}
The first case corresponds to a nonreciprocal model, namely a two-dimensional extension of the Hatano--Nelson model \cite{Hatano1996,Hatano1997,Gong2018}, obtained by setting
\begin{eqnarray}
\kappa_X^{(L)} &=& J_X \exp(h_X), \qquad \kappa_X^{(R)} = J_X \exp(-h_X), \\
\kappa_Y^{(L)} &=& J_Y \exp(h_Y), \qquad \kappa_Y^{(R)} = J_Y \exp(-h_Y),
\end{eqnarray}
where $J_X$, $J_Y$, $h_X$, and $h_Y$ are real parameters describing the horizontal/vertical hopping amplitudes and the associated imaginary gauge fields, respectively. This model exhibits the NHSE under OBC for essentially arbitrary boundary orientation $x$. As a representative example, let us consider $\theta=0$, i.e., a slab geometry with $x=X$ and $y=Y$. In this case one has $p=0$, $q=1$, and Eq.~(24) reduces to
\begin{eqnarray}
E \phi_n &=& J_Y \exp(h_Y)\,\phi_{n+1} + J_Y \exp(-h_Y)\,\phi_{n-1} \nonumber \\
&+& 2 J_X \cos \!\left( 2 \pi \alpha_x n + k_x a_X - i h_X \right) \phi_n.
\end{eqnarray}
\begin{figure*}[t]
 \centering
    \includegraphics[width=0.9\textwidth]{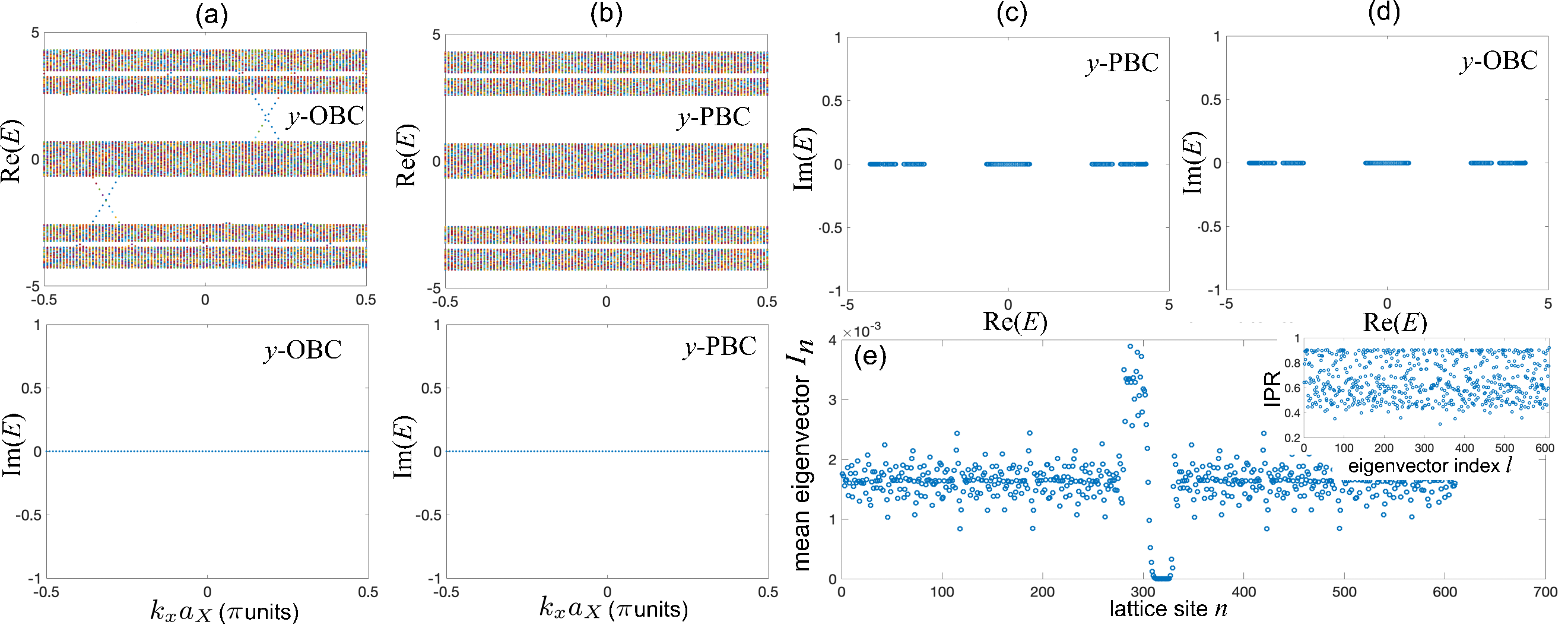}
   \caption{Same as Fig.2, but for an irrational magnetic flux $\Phi=(\sqrt{5}-1)/2$. Other parameter values are $J_X=2$, $J_Y=1$, $h_X=0$ and $h_Y=0.2$. In numerical simulations, the inverse of the golden ratio has been approximated by the rational number $\alpha \simeq 377/610$, i.e. ratio of Fibonacci numbers, and a lattice of size $L=610$ has been assumed. The energy spectrum is the same for $y$-PBC and $y$-OBC [apart for chiral edge states under $y$-OBC, visible in panel (a)], the eigenstates are bulk localized via Anderson localization and the NHSE is fully suppressed.}
    \label{fig4}
\end{figure*}
\begin{figure*}[t]
 \centering
    \includegraphics[width=0.8\textwidth]{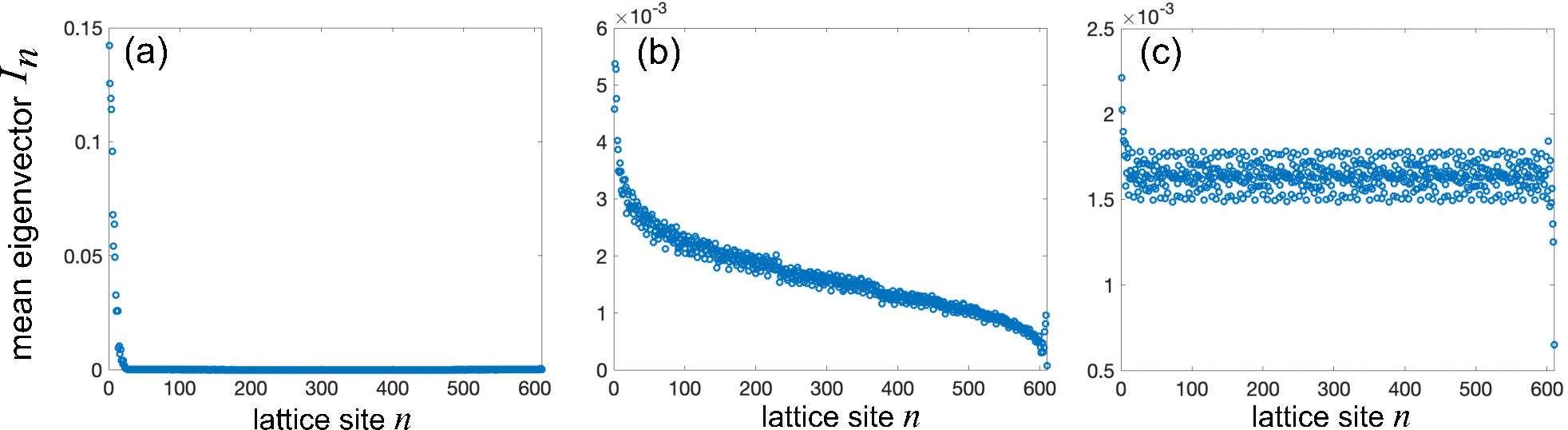}
   \caption{Suppression of the skin effect in the isotropic Harper-Hofstadter model ($J_X=J_Y=1$) under an irrational magnetic flux ($\Phi=(\sqrt{5}-1)/2$) by the application of non-reciprocal hopping parameter $h_X$  in the $X$ direction. The panels show the numerically-computed behavior of the mean eigenstate distribution $I_n$ of the Hamiltonian under $y-$OBC for $h_Y=0.2$, quasi-momentum $k_xa_X=0$, and for a few increasing values of $h_X$: (a) $h_X=0$, (b) $h_X=h_Y=0.2$, and (c) $h_X=0.3$.}
    \label{fig5}
\end{figure*}
where $\alpha_x=Ba_Xa_Y \equiv \Phi$ is the magnetic flux in each plaquette of the rectangular lattice. Equation~(32) represents a non-Hermitian extension of the Aubry--Andr\'e--Harper model, which has been the subject of extensive recent investigations \cite{AA0,AA1,AA2,AA3,AA4,AA5,AA6,AA7,AA8,AA8b,AA9,AA10,AA11}. In the absence of a magnetic field ($B=0$), Eq.~(32) reduces to the clean Hatano--Nelson model, which exhibits the NHSE under $y$-OBC for any $h_Y \neq 0$. When a magnetic field is present, one must distinguish between the cases of rational and irrational magnetic flux $\alpha_X$.

In the \emph{rational case}, Eq.~(32) describes a superlattice with nonreciprocal hopping. The corresponding energy spectrum under $y$-PBC consists of a set of closed loops in the complex plane, which differs substantially from the spectrum under $y$-OBC \cite{Longhi2023}---beyond the usual appearance of chiral edge states. As a result, the NHSE cannot be completely suppressed by the magnetic field for rational flux. An illustrative example is shown in Fig.~2 for $\Phi=\pi/2$. Figures~2(a) and 2(b) display the numerically computed energy spectrum (real and imaginary parts in the upper and lower panels, respectively) as a function of the quasi-momentum $k_x a_X$ in a lattice of size $L=500$, for $y$-OBC [Fig.~2(a)] and $y$-PBC [Fig.~2(b)] with parameter values $J_X=J_Y=1$, $h_X=0$, and $h_Y=0.2$. The spectra in the complex energy plane at $k_x a_X=0$ are shown in Figs.~2(c) and 2(d), corresponding to $y$-PBC and $y$-OBC, respectively. Notably, the $y$-PBC spectrum is distinct than the $y-$OBC spectrum and consists of closed loops that enclose a finite area, which is a clear signature of the NHSE. This feature is further confirmed in Fig.~2(e), where the numerically computed mean eigenvector distribution $I_n$ is plotted:
\begin{equation}
I_n = \frac{1}{L} \sum_{l=1}^L |\phi_n^{(l)}|^2,
\end{equation}
with $\phi_n^{(l)}$ denoting the normalized $l$-th eigenvector of Eq.~(32) under $y-$OBC. The inset shows the inverse participation ratio (IPR) of the $L$ eigenstates, defined as ${\rm IPR}_l = \sum_n |\phi_n^{(l)}|^4$. For extended states, the IPR takes small values scaling as $\sim 1/L$, while for localized states it remains finite and independent of system size $L$. As one can see, the eigenvectors are localized and preferentially accumulate near the left lattice edge, indicating the persistence of skin localization. However, for small magnetic flux $\Phi$, the skin topological area---namely, the area enclosed by the closed spectral loops---shrinks, and the eigenstates show a tendency toward bulk localization. This crossover is associated with bulk Landau-type localization in the low-energy regime of the model \cite{M1,M9} which competes with skin localization, as illustrated in Fig.~3. \par
In the \emph{irrational case}, the eigenstates of Eq.~(32) under $y$-PBC undergo a localization--delocalization transition at the critical point   
\begin{equation}
|h_Y| + \ln \left| \frac{J_Y}{J_X} \right| - |h_X| = 0
\end{equation}
which has been derived in previous works (see e.g. \cite{AA8}). In fact, in the Hermitian limit $h_X=h_Y=0$ Eq.~(32) reduces to the usual Aubry-Andr\'e model, which displays a localization-delocalization transition at $|J_X|=|J_Y|$ with a Lyapunov exponent (inverse of localization length) given by $\lambda=\ln |J_X/J_Y|$ in the localized phase $|J_X/J_Y|>1$. The addition of non-reciprocal hopping via the imaginary gauge field $h_Y$ tends to delocalize the eigenstates, i.e. reduce the Lypaunov exponent, while the imaginary phase $h_X$ in the incommensurate potential tends to localize the eigenstates, i.e. to increase the Lyapunov exponent. Overall, the critical point of the transition, corresponding to vanishing of the Lyapunov exponent, is provided by Eq.(34). 
In particular, bulk Anderson localization occurs whenever  
\begin{equation}
|h_Y| + \ln \left| \frac{J_Y}{J_X} \right| - |h_X| < 0,
\end{equation}
indicating that in this regime the NHSE is completely suppressed by the magnetic field and the eigenstates become localized in the bulk. As a representative example, consider reciprocal hopping along the $X$ direction ($h_X=0$), where the incommensurate potential in Eq.(32) is real. In this case, the above condition shows that magnetic flux can eliminate the NHSE provided $J_X > J_Y$ and the nonreciprocal hopping parameter in the $Y$ direction satisfies $|h_Y| < \ln |J_X/J_Y|$.  
An illustrative example of magnetic suppression of skin localization in this regime is presented in Fig.~4, where the magnetic flux $\Phi=(\sqrt{5}-1)/2$ is the inverse golden mean and the parameter values satisfy Eq.~(35). As evidenced by the behavior of the IPR [inset of Fig.~4(e)], all eigenstates under $y$-OBC are localized. However, unlike the skin modes, they are no longer confined to the boundary; instead, they are localized in the bulk at various lattice sites, yielding an almost uniform mean eigenvector distribution [Fig.~4(e)].

A special case is provided by the isotropic Harper--Hofstadter model with $J_X = J_Y$. The preceding analysis indicates that the NHSE cannot be suppressed by the magnetic field when $h_X = 0$. Indeed, in the Hermitian limit ($h_Y=0$), Eq.~(32) reduces to the Harper equation, whose eigenstates are critical, being neither exponentially localized nor fully extended. As soon as a small nonreciprocity $h_Y \neq 0$ is applied to the vertical hopping rates, all states become skin modes localized at one edge under $y$-OBC.  
However, increasing $h_X$, i.e., introducing nonreciprocity along the $x$ direction with PBC, allows condition~(35) to be satisfied, and the NHSE can then be suppressed by the magnetic field. Remarkably, nonreciprocity along the periodic $x$ direction can inhibit the emergence of skin localization along the transverse $y$ direction with OBC. This counterintuitive effect is illustrated in Fig.~5.

\begin{figure*}[t]
 \centering
    \includegraphics[width=0.9\textwidth]{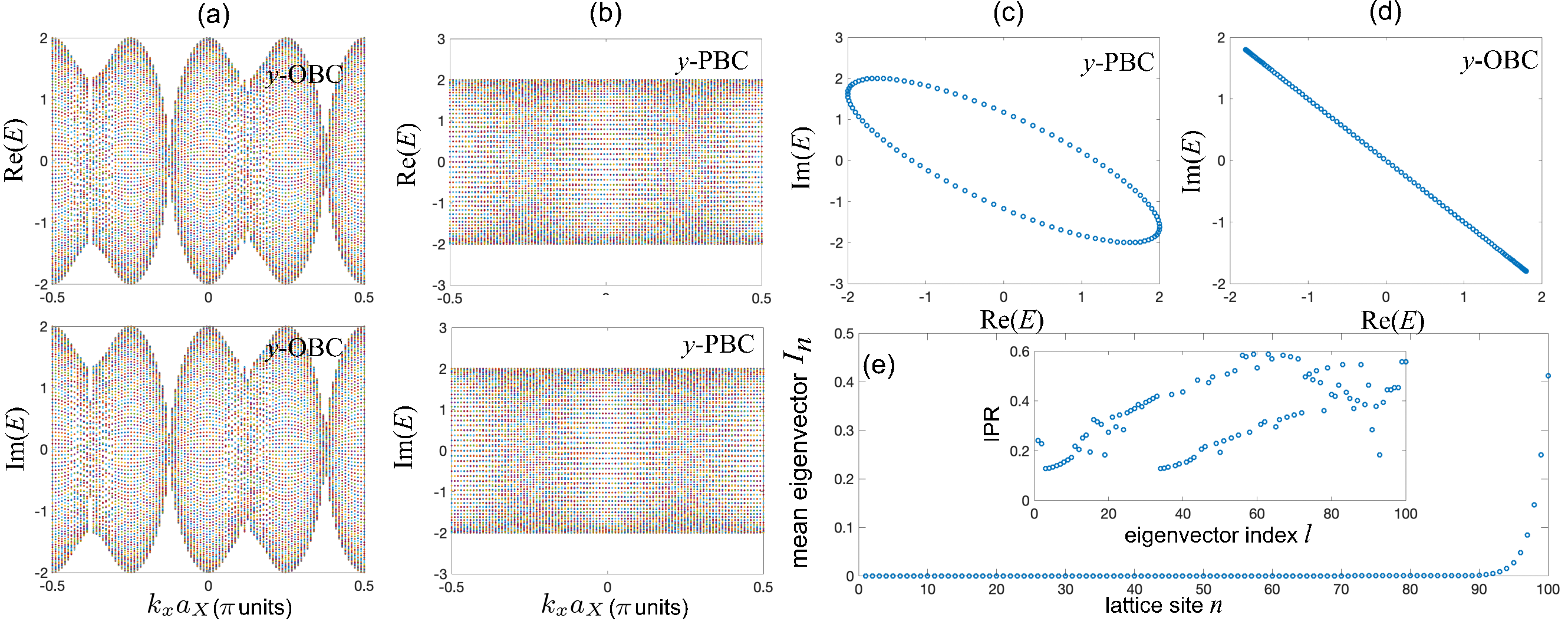}
   \caption{Geometry-dependent NHSE in the reciprocal Harper-Hofstadter model in a slab geometry. (a,b) Energy spectrum versus quasi-momentum $k_xa$ under $y$-OBC [panels (a)] and $y$-PBC [panels (b)], for $a_X=a_Y=\sqrt{2} a$, corresponding to $\theta= \pi/4$ and $\alpha_x=\alpha_y=\Phi /2$, and for parameter values $\kappa_X=1$, $\kappa_Y=i$ and $\Phi=0$. Lattice size in the $x$ direction is $L=100$. Real and imaginary parts of $E$ are shown in the upper and lower panels, respectively. (c,d) Energy spectrum in complex energy plane for $k_xa=2 \pi/5$ and for $y$-PBC [panel (c)] and $y$-OBC [panel (d)]. (e) Mean eigenvector distribution $I_n$ under $y$-OBC, clearly showing the presence of the NHSE. The inset in (e) shows the IPR of the $L$ eigenstates under $y-$OBC.}
    \label{fig2}
\end{figure*}

\begin{figure*}[t]
 \centering
    \includegraphics[width=0.9\textwidth]{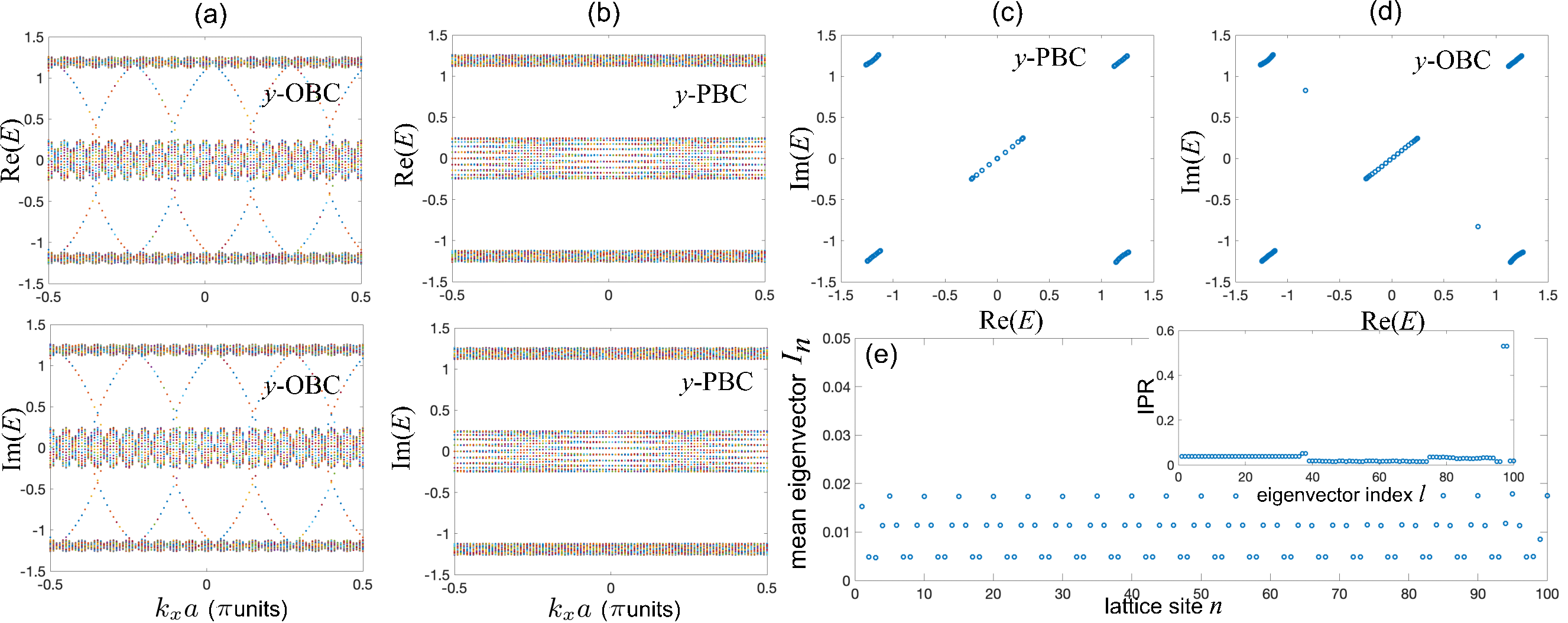}
   \caption{Suppression of the geometry-dependent NHSE in the reciprocal Harper-Hofstadter model under a weak magnetic field. Parameter values are as Fig.6 except for a magnetic flux $ \Phi=1/5$.}
\end{figure*}

\subsection{Reciprocal lattices}
The second case corresponds to non-Hermitian lattices with reciprocal hopping, $\kappa_X^{(L)}=\kappa_X^{(R)} \equiv \kappa_X$ and $\kappa_Y^{(L)}=\kappa_Y^{(R)} \equiv \kappa_Y$, which display the geometry-dependent NHSE \cite{R21} in the absence of the magnetic field. In fact, for $B=0$ the dispersion curve of the lattice band reads
\begin{equation}
E(k_X,k_Y)=2 \kappa_X \cos(k_Xa)+2 \kappa_Y \cos(k_Ya_Y)
\end{equation} 
\begin{figure*}[t]
 \centering
    \includegraphics[width=0.98\textwidth]{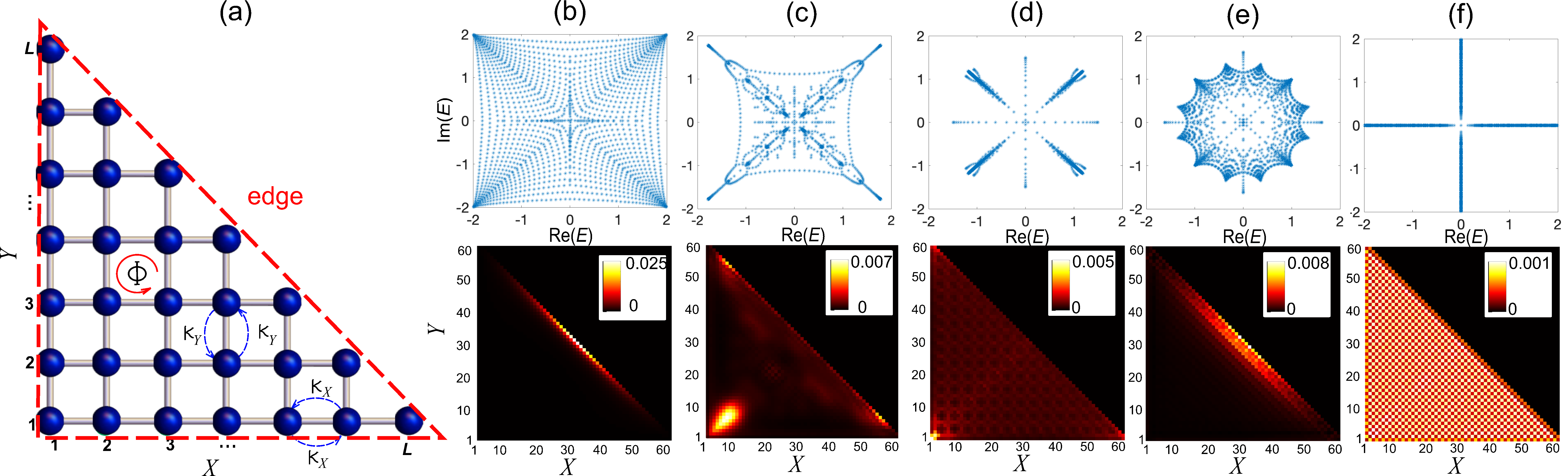}
   \caption{ (a) Schematic of a 2D square lattice ($a_X=a_Y$) with reciprocal hopping amplitudes $\kappa_X$ in the horizontal $X$ direction and $\kappa_Y$  in the vertical $Y$ direction. OBC are assumed along a triangular contour (dashed red curve). The total number of sites in the lattice with OBC is $N=L(L+1)/2$. Parameter values used in the numerical simulations are $\kappa_X=1$, $\kappa_Y=i$ and $L=60$. (b-f) Energy spectrum (upper panels) and mean eigenvector distribution $I_{X,Y}$ on a pseucolor map (lower plots) for a few increasing values of the magnetic flux $\Phi=a_Xa_YB$: (b) $\Phi=0$;  (c) $\Phi=\pi/20$;
 (d) $\Phi=\pi/4$; (e) $\Phi=\pi/3$; (f) $\Phi=\pi/2$.}
\end{figure*}
which displays in complex $E$ plane a non-vanishing spectral area for generic non-Hermitian (yet reciprocal) hopping amplitudes $\kappa_X$, $\kappa_Y$. For an edge along a rather arbitrary $y$ direction, one expects skin localization of bulk modes toward the edge, with the exception when $y$ is one of the two principal directions $X$ or $Y$ of the Bravais lattice, i.e. at the cut angles $\theta=0, \pi/2$. In the following analysis, we assume a cut angle $\theta$ satisfying Eq.(23) with $p=q=1$, i.e. along the main diagonal of the rectangle forming the primitive cell of the lattice.  In this case, the eigenvalue equation (15) takes the form
\begin{equation}
E \phi_n=W^{(L)}_{n} \phi_{n+1}+W^{(R)}_{n-1} \phi_{n-1}
\end{equation}
where we have set
\begin{widetext}
\begin{eqnarray}
W_n^{(L)} & = & \kappa_X \exp \left( -2 \pi i \alpha_xn-ik_xa_X \cos \theta - i\pi \Phi \cos^2 \theta  \right)+ \kappa_Y \exp \left( 2 \pi i \alpha_y n+ik_xa_Y \sin \theta + i \pi \Phi \sin^2 \theta  \right) \\
W_n^{(R)} & = & \kappa_X \exp \left( 2 \pi i \alpha_xn+ik_xa_X \cos \theta + i \pi \Phi \cos^2 \theta  \right)+ \kappa_Y \exp \left( -2 \pi i \alpha_y n-ik_xa_Y \sin \theta -i  \pi \Phi \sin^2 \theta  \right),
\end{eqnarray}
\end{widetext}
and $\alpha_x= \Phi \cos^2 \theta$, $\alpha_y=\Phi \sin^2 \theta$ and $\Phi= B a_X a_Y$. 
Therefore, in the effective 1D lattice model [Eq.(37)] the magnetic field introduces a spatial modulation of the effective left and right hopping rates, $W_{n}^{(L,R)}$.
In the absence of the magnetic field ($B=0$), the lattice is homogeneous with hopping rates 
\begin{eqnarray}
W^{(L)} & = & \kappa_X \exp \left(-ik_xa_X \cos \theta \right)+ \kappa_Y \exp \left( ik_xa_Y \sin \theta \right) \;\; \nonumber \\
W^{(R)} & = & \kappa_X \exp \left( ik_xa_X \cos \theta  \right)+ \kappa_Y \exp \left( -ik_xa_Y \sin \theta \right). \;\; \nonumber
\end{eqnarray}
In a non-Hermitian yet reciprocal lattice, i.e. when $\kappa_{X} \neq \kappa_X^*$ and/or $\kappa_{Y} \neq \kappa_Y^*$, for rather arbitrary quasi-momentum $k_x$ one has 
$|W^{(L)}| \neq |W^{(R)}|$, indicating the appearance of the NHSE. This is shown, as an example, in Fig.6. When the magnetic field is turned on, the hopping rates are modulated, with $|W_n^{(L)}| > |W_n^{(R)}|$ in some sites and $|W_n^{(L)}| < |W_n^{(R)}|$ in the other sites. Specifically, one has
\begin{equation}
 \frac{|W_n^{(R)}(E)|^2}{ | W_n^{(L)}(E)|^2}=
 \frac{ |\kappa_X|^2+|\kappa_Y|^2+ 2 |\kappa_X| |\kappa_Y| \cos (\omega_n+\rho)}  { |\kappa_X|^2+|\kappa_Y|^2+ 2 |\kappa_X| |\kappa_Y| \cos (\omega_n-\rho)} 
\end{equation}
where we have set $\omega_n= \pi \Phi (2n+1)+k_x a/( \sin \theta \cos \theta)$ and $\rho$ is the phase of $\kappa_X \kappa_Y^*$. This relation indicates that, while the model (37) is locally non-reciprocal with unbalanced left/right hopping rates  for $\rho \neq 0, \pi$, i.e. $|W_n^{(R)} / W_n^{(L)}| \neq 1$, the modulation induced by the magnetic field tends to restore {\em global reciprocity} \cite{R30c,Midya2024} favoring the mitigation or even the suppression of the NHSE. 
For example, if the magnetic flux in each plaquette $ \Phi= B a_X a_Y$ is chosen such that $\Phi= \pi/2$, then $\omega_n$ mod $2 \pi$ takes only two opposite values and thus the ratio in Eq.(40) alternates between two values $\lambda$ and $1 / \lambda$, indicating perfect balance between left and right hopping on average and suggesting full suppression of the NHSE. More generally, as demonstrated in the Appendix A a small magnetic flux, or an irrational magnetic flux, 
can suppress the skin effect. An illustrative example of magnetic suppression of geometry-dependent skin effect for a weak magnetic flux is shown in Fig.7.

\subsection{Discussion}  

It should be emphasized that the present analysis has been carried out in a strip geometry invariant along the $x$ direction [Fig.~1(a)] and in Landau gauge, where periodic boundary conditions allow for an analytic reduction to an effective one-dimensional model. This framework provides clear insight into the distinct mechanisms by which magnetic fields control skin localization. Nevertheless, introducing boundaries along the $x$ direction, or more generally considering polygonal geometries, can modify the routes to boundary localization, potentially giving rise to edge- or corner-localized modes controlled by flux.  
As an illustrative example, consider the reciprocal model of Sec.~IV.B with a square lattice ($a_X=a_Y$, $\theta=\pi/2$) terminated by a triangular boundary, as shown in Fig.~8(a). The numerically computed mean eigenvector distribution $I_{X,Y}$ and the corresponding energy spectrum for increasing values of the magnetic flux $\Phi=B a_X a_Y$ are displayed in Figs.~8(b-f). For vanishing flux, skin localization is observed along the diagonal edge of the triangle, a manifestation of the geometry-dependent skin effect \cite{R21} [Fig.~8(b)]. For a weak magnetic flux,  the boundary accumulation is mitigated and replaced by a tendency toward bulk and corner localization [Figs.~8(c-d)]. As the magnetic flux is increased, according to the analysis of Sec.~IV.B and Appendix A, the suppression of the skin effect depends sensitively on the value of the magnetic flux $\Phi$, since global reciprocity is only partially restored. For instance, at $\Phi=\pi/3$ the suppression remains incomplete [Fig.~8(d)], while at $\Phi=\pi/2$ the NHSE is fully suppressed [Fig.~8(f)]. These results exemplify how boundary geometry and magnetic fields can cooperate to reshape localization patterns beyond the strip geometry.  
While a general universal theory for arbitrary polygonal boundaries remains an open challenge, the triangular-lattice example highlights the versatility of magnetic fields as a means of tailoring boundary phenomena in non-Hermitian systems. It also provides a natural bridge to the broader implications and perspectives discussed in the Conclusions.

\section{Conclusions}
In this work, we have presented a theoretical analysis of magnetic control of the non-Hermitian skin effect in two-dimensional single-band lattices with strip geometries. By examining a non-Hermitian extension of the anisotropic Harper--Hofstadter model, we clarified the respective roles of flux, nonreciprocity, and boundary geometry. Our results show that magnetic fields suppress the geometry-dependent NHSE in reciprocal systems, rendering it fragile even under weak flux, whereas skin localization in nonreciprocal systems is more robust but can nevertheless be mitigated or suppressed through gauge-field-induced bulk localization mechanisms. Two distinct routes by which magnetic fields influence boundary accumulation have been identified: restoration of effective reciprocity in reciprocal models and Landau- or Anderson-type localization in nonreciprocal ones.

These findings shed light on the mechanisms underlying the interplay between gauge fields and non-Hermitian topology, helping to clarify how magnetic flux affects boundary accumulation in different regimes. While our analysis focuses on a representative single-band model, the general trends identified here may be relevant to a broader class of non-Hermitian systems.
Several directions for further study emerge. Extensions to multiband systems, higher-order skin effects, or models with spin--orbit coupling may reveal additional mechanisms of magnetic control. The interplay between gauge fields and many-body or nonlinear non-Hermitian systems represents another timely challenge, with potential to uncover new forms of collective localization.

\appendix
\section{Magnetic suppression of geometry-dependent skin effect}
In this Appendix it is shown that application of a weak magnetic field, or more generally of an irrational magnetic flux, to the reciprocal non-Hermitian model of Sec.IV.B suppresses the geometry-dependent skin effect.  To this aim, let us consider the eigenvalue equation (37), given in the main text
\begin{equation}
E \phi_n=W_n^{(L)} \phi_{n+1}+W_{n-1}^{(R)} \phi_{n-1}
\end{equation} 
and assume OBC, i.e. $\phi_0=\phi_{L+1}=0$, where $L$ is the lattice size. 
As discussed in the main text, an extensive NHSE arises in the absence of the magnetic flux for a rather generic value of the quasi-momentum $k_x$, i.e. there is a macroscopic piling of  the eigenstates $\phi_n$ to one boundary. To prove that a non-vanishing magnetic field can suppresses the NHSE, let us first introduce the gauge transformation
\begin{equation}
\phi_n= \xi_n \exp(i \pi \alpha_x n^2+i \sigma_xn-i \pi \alpha_x n)
\end{equation} 
where we have set $\sigma_x \equiv k_xa_X \cos \theta+ \pi \Phi \cos^2 \theta$. Then, using Eqs.(38) and (39),
the eigenvalue equation (A1) takes the form
\begin{equation}
E \xi_n=\tilde{W}_n^{(L)} \xi_{n+1}+\tilde{W}_{n-1}^{(R)} \xi_{n-1}
\end{equation} 
with the OBC  $\xi_0=\xi_{L+1}=0$, where we have set
\begin{equation}
\tilde{W}_n^{(L)}= \kappa_X+\kappa_Y \exp(i \omega_n) \; , \;  \tilde{W}_n^{(R)}= \kappa_X+\kappa_Y \exp(-i \omega_n) 
\end{equation}
and 
\begin{equation}
\omega_n=\pi  \Phi (2n+1)+ \frac{k_x a}{\sin \theta \cos \theta}.
\end{equation}
Let us assume that the magnetic flux in each plaquette,  $\Phi= B a_X a_Y$, is  weak or close to an irrational number, such that there exists a {\em large integer} $q$ such that $\tilde{W}^{(L,R)}_{n+q} \simeq \tilde{ W}^{(L,R)}_{n}$. Let us then  write Eq.(A3) in transfer-matrix form
\begin{equation}
\left(
\begin{array}{c}
\xi_{n+1} \\
\xi_n
\end{array}
\right)= \mathcal{M}_n(E) \left(
\begin{array}{c}
\xi_{n} \\
\xi_{n-1}
\end{array}
\right)
\end{equation}
with transfer matrix
\begin{equation}
\mathcal{M}_n(E)=\left(
\begin{array}{cc}
\frac{E}{\tilde{W}_n^{(L)}} & -\frac{\tilde{W}_{n-1}^{(R)}}{\tilde{W}_n^{(L)}} \\
1 & 0
\end{array}
\right).
\end{equation}
Over $Nq$ sites one has
\begin{equation}
\left(
\begin{array}{c}
\xi_{Nq+1} \\
\xi_{Nq}
\end{array}
\right)= \mathcal{S}^N (E)
\left(
\begin{array}{c}
\xi_1 \\
 \xi_{0}
\end{array}
\right)
\end{equation}
where we have set
\begin{equation}
\mathcal{S}(E)= \mathcal{M}_q (E) \mathcal{M}_{q-1} (E)... \mathcal{M}_1(E).
\end{equation}
Under OBC in a lattice comprising $L=Nq$ sites, $\xi_{0}=\xi_{Nq+1}=0$ and thus from Eq.(A) one obtains
\begin{equation}
\left( \mathcal{S}^{N} (E)  \right)_{1,1}=0
\end{equation}
and
\begin{equation}
\xi_{Nq}= \left( \mathcal{S}^{N} (E) \right)_{2,1} \xi_1.
\end{equation}
The first condition, Eq.(A10), determines the  set of $Nq$ allowed energies  $E$ (energy spectrum), whereas Eq.(A11) indicates that the ratio $| \phi_{Nq} / \phi_1|=| \xi_{Nq} / \xi_1|$ scales with $N$ as 
$ \left| \left( \mathcal{S}^{N} (E) \right)_{2,1} \right|$. The NHSE arises whenever for almost all energies $E$ in the spectrum $\left( \mathcal{S}^{N} (E) \right)_{2,1}$ diverges or vanishes exponentially as $N \rightarrow \infty$. Indicating by $\lambda_{1,2}(E)$ the two eigenvalues of the matrix $\mathcal{S}(E)$ and by $\mathcal{T}$ the matrix of eigenvectors, taking into account that
\begin{equation}
\mathcal{S}^N(E)=\mathcal{T} \left( 
\begin{array}{cc}
\lambda_1^N(E) & 0 \\
0 & \lambda_2^N (E)
\end{array}
\right)
\mathcal{T}^{-1}
\end{equation}
the condition (A10) implies that, when $E$ is in the spectrum, then $|\lambda_1(E)|=|\lambda_2(E)|$ for almost any energy $E$, whereas Eq.(A11) shows that the NHSE arises whenever at least one of the two eigenvalues has modulus different than one for almost any energy $E$ in the spectrum. As we prove below, for any $E$ one has $| \det \mathcal{S}(E)|=| \lambda_1(E) \lambda_2(E)|=1$, and since $|\lambda_1(E)|=|\lambda_2(E)|$ for almost any energy in the spectrum, one concludes $| \lambda_1(E)|=|\lambda_2(E)|=1$ for almost any energy in the spectrum and thus  the NHSE effect is absent.\\ 
We now prove that $|\det \mathcal{S}(E)|=1$. To this aim, let us notice that
\begin{eqnarray}
| \det \mathcal{S}(E)|^2 & = &  | \det \mathcal{M}_q |^2 | \det \mathcal{M}_{q-1}|^2 ... |\det \mathcal{M}_1|^2 = \nonumber \\ 
& = & \frac{\prod_{n=1}^q |\tilde{W}_n^{(R)}(E)|^2}{\prod_{n=1}^q | \tilde{W}_n^{(L)}(E)|^2} = \\
& = & \frac{\prod_{n=1}^q |W_n^{(R)}(E)|^2}{\prod_{n=1}^q | W_n^{(L)}(E)|^2} \nonumber
\end{eqnarray}
where we used the cyclic property $W_{n}^{(L,R)}(E)=W_{n+q}^{(L,R)}(E)$. Physically, the condition $|\det \mathcal{S}(E)|=1$  corresponds to global reciprocity of the system. From the explicit form of $\tilde{W}_{n}^{(L,R)}(E)$ given by Eq.(A4), one obtains
\begin{equation}
| \det \mathcal{S}(E)|^2  =    \frac{\prod_{n=1}^q |\kappa_X|^2+|\kappa_Y|^2+ 2 |\kappa_X| |\kappa_Y| \cos (\omega_n+\rho)}  {\prod_{n=1}^q |\kappa_X|^2+|\kappa_Y|^2+ 2 |\kappa_X| |\kappa_Y| \cos (\omega_n-\rho)} 
\end{equation}
where $\rho$ is the phase of $\kappa_X \kappa_Y^*$. From Eq.(A14) one obtains
\begin{eqnarray}
\ln | \det \mathcal{S}(E)|^2 = \nonumber \\ 
   \sum_{n=1}^q \ln \left\{ |\kappa_X|^2+|\kappa_Y|^2+ 2 |\kappa_X| |\kappa_Y| \cos (\omega_n+\rho) \right\} \nonumber \\
 -   \sum_{n=1}^q \ln \left\{ |\kappa_X|^2+|\kappa_Y|^2+ 2 |\kappa_X| |\kappa_Y| \cos (\omega_n-\rho) \right\}. \;\;\;
\end{eqnarray}
In the large $q$ limit, as $n$ varies from $1$ to $q$ the phase $\omega_n$ mod $ 2 \pi$ uniformly fills the interval $(0, 2 \pi)$, so that one can write
\begin{eqnarray}
\sum_{n=1}^q \ln \left\{ |\kappa_X|^2+|\kappa_Y|^2+ 2 |\kappa_X| |\kappa_Y| \cos (\omega_n \pm \rho) \right\}  & \simeq  \nonumber \\
\frac{q}{2 \pi} \int_0^{2 \pi} d \omega \ln \left\{ |\kappa_X|^2+|\kappa_Y|^2+ 2 |\kappa_X| |\kappa_Y| \cos (\omega \pm \rho) \right\} \nonumber \\
\end{eqnarray}
Clearly, the integral on the right hand side of Eq.(A16) is independent of $\rho$, and thus the two sums on the right hand side of Eq.(A15) are the same, yielding $\ln | \det \mathcal{S}(E)|^2=0$, i.e. $ | \det \mathcal{S}(E)|=1$.\\
Finally, we mention that for rational flux $\Phi=p/q$ with relatively small $q$  the condition $| \det \mathcal{S}(E)|=1$ cannot be rather generally strictly satisfied, i.e. global reciprocity restoration is imperfect, and the skin effect is only partly mitigated. On the other hand, in some special conditions one exactly has $| \det \mathcal{S}(E)|=1$ even for small $q$, for example for a flux $\Phi=p/q=1/2$, one has exactly $| \det \mathcal{S}(E)|=1$, leading to perfect washing out of the NHSE.

\end{document}